# High Reliable secure data collection using Complexity exchanging code key method addressing protocol in Wireless Sensor Network

Dr.V.jayaraj[1], M.Indhumathi[2]
[1] Associate Professor,Department of Computer science, Kajamalai Campus,
Bharathidasan University, Tiruchirappalli, India

[2] Research Scholar ,Department of Computer Science, Kajamalai Campus,
Bharathidasan University, Tiruchirappalli, India
[1]jaya_v2000@yahoo.com, [2]umindhu@gmail.com

**Abstract**:  A Wireless Sensor Network (WSN) is emerging field in Information and communication technology. In WSN data transmission and data collection are unsecure because of sensor node Incompatibility. So providing security to Sensor network is very important. The key based Mechanism is secure data collection and it's mainly used to guarantee data confidentiality. Range pair wise key is largely used due to the necessary of data encryption and decryption between each pair range of communication node. Fixed key mechanism difficult for the attacker to detect the regularity of the randomly generated key chain function in privacy homomorphism (PH).PH means no intermediate node to encrypt and decrypt only direct collect and aggregate data for encryption and decryption. It is special key based scheme. It's totally based on beta distribution and some statistical tool using key code method by using Complexity exchanging code key method addressing protocol. We show how to reduce significant attacks and secure data collection on wireless sensor network.

# 1      Introduction

A wireless sensor network (WSN) is a wireless network consisting of spatially
distributed autonomous devices that use sensors to monitor physical or environmental
conditions. These autonomous devices, or nodes, combine with routers and a gateway
to create a typical WSN system. The distributed measurement nodes communicate
wirelessly to a central gateway, which provides a connection to the wired world
where you can collect, process, analyze, and present your measurement data. To
extend distance and reliability in a wireless sensor network, you can use routers to
gain an additional communication link between end nodes and the gateway. The
resource limitations of WSNs, the data may be tampered easily during transmission. It
may happen that data reach incorrect destinations or received incorrect data. It is an
important issue for WSNs that base stations must receive correct data and ensure the
data confidentiality.

The inherent limitations of WSNs, eavesdropping or jamming may easily occur, and
WSNs are often applied to abominable environments and conditions, such as military
affairs and the fire emergencies. Therefore, security has become an important issue in
the world of WSNs. The main functionalities of WSN are sensing data and send back
to base station. Therefore, it is an important issue to make sure that data can reach the
base station correctly and efficiently. Therefore, we need to pay special attention to
security issues when data is being forwarded.  The data dissemination service is that
each node transfers the received packets to another node until the packets, following
the routes selected by the routing protocol, reach the destination.

Therefore, it is very important to establish a secure routing path, a correct and reliable
routing path to ensure that data can reach the correct destination and to maintain the
confidentiality and integrity of the data packet. Main security threats in WSN are: 1)
Radio links are insecure – eavesdropping / injecting faulty information is possible. 2)
Sensor nodes are not temper resistant – if it is compromised attacker obtains all
security information The various Attacker types in wireless sensor network are *Mote-
class:* attacker has access to some number of nodes with similar characteristics /
laptop-class: attacker has access to more powerful devices. *Outside / inside:* attacker
compromised some number of nodes in the network.

In WSN Key management protocol must establish a key between all sensor nodes that
must exchange data securely and also it facilitates the node addition / deletion should
be supported. It should work in undefined deployment environment; unauthorized
nodes should not be allowed to establish communication with network nodes.

The various key management approaches the already proposed are *Pre-deployed keying:* Straightforward approaches, Eschenauer / Gligor random key pre-deployment, Chan / Perrig q-composite approach, Zhu / Xu approach, DiPietro smart attacker model and PRK protocol; *Key derivation information pre-deployment:* Liu / Ning polynomial pre-deployment; *Self-enforcing autonomous approaches:* Pair-wise asymmetric (public key).

Key management poses a main concern for security operation in sensor network. Many key management protocols are proposed homogeneous sensor network, however these networks have limited performance and security heterogeneous sensor network are proposed to outcome these drawbacks. In this method using wireless sensor network that consisted of three types of key method, that's are Random key, deterministic key and Hybrid key. Random key can randomly chooses several key from the key pool and to create chain. Deterministic key can use dynamic computation to generate key that can enhance the connection between sensor nodes. Key pre distribution is the method of distribution of key on to nodes before deployment. The nodes build up the network using secret key after deployment. When the reach their target position. Key predistribution schemes are various methods has been developed by academicians for a better maintenance of key management. A key predistribution has three phases key distribution, shared key, discovery, path key establishment

## 1.1  Security attacks on wireless sensor network:

***The lists of attack faced by a wireless sensor network are:***
Node outage, Physical attacks, Message corruption, False node, Node replication attacks, Passive information gathering Attacks against privacy, Monitor eves dropping, Traffic analysis, Camoutflages, Adversaries, Denial of service, Node subversion, Node malfunction, Routing attacks, Selective forwarding, Sinkhole, Wormhole.

## 1.2  Role of Key Management and Agreement Protocols

***The various Key concepts are:***
***Key:*** Symmetric key which is used to secure communication among two or more sensor nodes.
***Keychain:*** List of keys or keying materials which are stored on a sensor node.
***Key pool:*** List of all keys or keying materials which are used in the WSN.
***Link key:*** This is used to secure communication over a direct wireless link.

***Path key:*** This is used to secure communication over multihop wireless link, through one or more  sensor node.

***Pair wise:*** It's used secure unicast communication between  pair of sensor nodes.

***Forward secrecy:*** It requires that a departed/expelled member of a group should ensure that the departed/expelled member cannot decrypt group data after it leaves the group.

***Backward secrecy:*** It requires that a new member should not have access to the previous (old) group key. This ensures that the newly joined node cannot decrypt messages that are exchanged in the group before the node joined the group access to future keys after it has left the group cannot decrypt group data after it leaves the group.

***Rekeying:*** Rekeying refers to the processing of changing the group key securely upon a membership change. Rekeying should be triggered by the protocol after each membership change to ensure forward and backward secrecy.

***The Authentication key methods are:***
   **KDC** - Key Distribution Centre
   **HWSN** - Hierarchical Key Chain WSN
   **MAC**- Message authentication   node
   **PRF**-Pseudo Random   Function
   **ENC** - Encryption.
   **DAG** - Directed   Acyclic Graph.

The key management and key agreement protocols should cope with the demands of various applications. Besides confidentiality, integrity, and authenticity.

## 2      Secure Data Collection

   The key based Mechanism is secure data collection in mainly used to guarantee data confidentiality range pair wise key is largely used due to the necessary of data encryption and decryption between each pair range of communication node. Fixed key mechanism difficult for the attacker to detect the regularity of the randomly generated key chain function in Privacy Homomorphism (PH).PH means no intermediate node to encrypt and decrypt, only direct collect and aggregate data. It is special key based scheme.

It is totally based on beta distribution and some statistical using key code method we show how to reduce significant attacks and secure data collection on wireless sensor network.

## 2.1   Data Exchange Protocol

Before the sensor are deployed in network (each sensor x is supplied with flowing item). N is the number of sensor thus  the total number of key that need to be in sensor  network (n+1)/2 sink node key SI.

***Step1:*** Total number of sensor data allocated in the network $x_n$

$$A=f(x_1, x_2, x_3 \ldots \ldots .. x_n)$$

***Step2:*** Sink node only allocated tempravary key in all sensor  node.

$x_1=2$ (Temp key(2))

$x_2=5$ (Temp key(5))

.

.

.

$x_n = n$ (Temp key (n))

***Step3:*** Where sink node SI to store the original key because sink node only maintained in secret key.

$S_l \rightarrow x_1= 6 \; \rightarrow$ (org  K(6))

$S_l \leftarrow x_1 = 7 \rightarrow$ (org K(7))

…

…

. $X_n=12(org(12))$

***Step4:*** After get the original Information every sink node is to change the temp key and also original key.

$S_l$: verify($x_1, x_2$)

## 2.2   Permutation boxes and compression box methods

### P-Box

Here we define the permutation boxes (P – Box) for mapping the possible keys, the straight P – Boxes have the n$\rightarrow$possible inputs and m$\rightarrow$ possible outputs. Here for example 6 possible mapping of 3X3 P – Boxes.

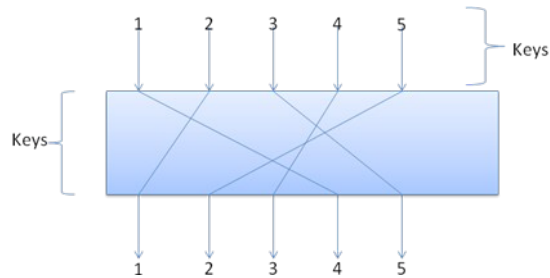

Fig 1: P – Box key

**Compression P-Box**

Here we introduce the compression method for the above P – Box methods it's called as Compression P – Box.  A Compression P – Box have an n → inputs and m → outputs in after the compression of possible keys. Were Compressed P – Box is defend against eavesdropping and also it's cannot defend against impersonation. The diagrammatic prototype is described below:

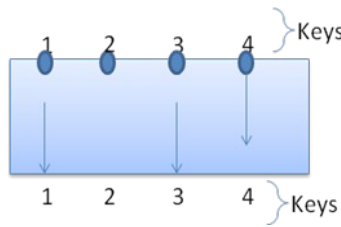

Fig 2: A compressed P – Box key

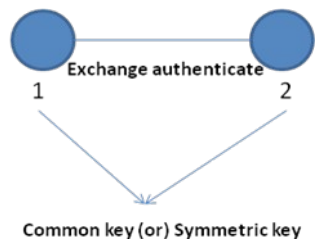

Fig 3: Defend against eavesdropping

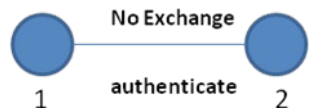

Fig 4: Can't defend against impersonation

### 2.3   Mutual authentication

In the mutual authentication the Sensor **X** authenticate sensor **Y** and Sensor **Y** authenticate sensor **X.** Encrypt and later decrypt all Exchange data message between **X** and **Y.**  The prototype of mutual authentication is represented below:

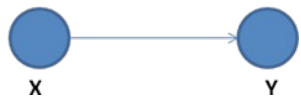

Fig 5: **X** authenticate **Y**

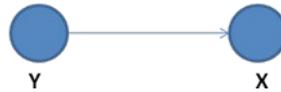

Fig 6: **Y** authenticate **X**

Here X is adjacent Y means to get the Y information. In some time the mutual authentication is adversary to other unknown node. Example the Z adjacent to X or Z adjacent to Y. Here Z copy the both information from X and Y.

*Note:* (i). X adjacent Y (neighbored). (ii)Two sensor could be stationary. This model is described as below:

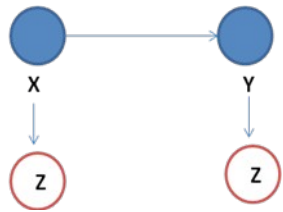

Fig 7: Here Z is the adversary Node in Mutual authentication.

## 3     Communication Device

Communication  device is used  to  exchange data between   individual node by transiver, single hop and multihop . This task is to  convert  a  bit  stream coming from  a  micro controller  and  convert  them  to  radio  waves

***Transiver:***

It's convert   week  signal  to  change  strong  signal  amplifier.

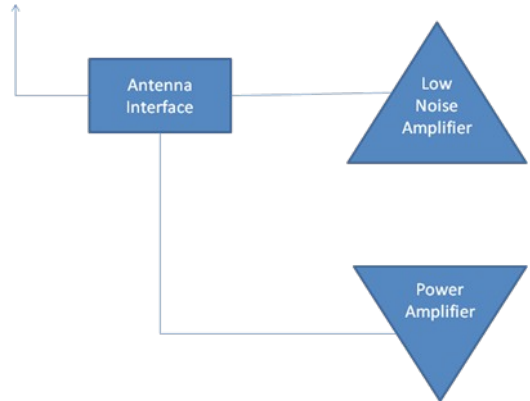

Fig 8: Transiver

***Single hop:***
Were single hop is power limitation and its limitation distance between sender and receiver. Its provide the direct communication between source and destination. Sink node is not always possible in single hop.

***Multi hop:***
In the multi hop direct communication is impossible because of distance (or) Obstacles.

**3.1          Path loss attenuation**

Distance-dependent loss of power, called **path loss**. The received power at a distance of $d \geq d_0$ between transmitter and receiver is described by the **Friis free-space equation.**

$$P_{rcvd}(d) = \frac{P_{tx}.G_t.G_r.\lambda^2}{(4\pi)^2.d^2.L}$$

$$\frac{P_{tx}.G_t.G_r.\lambda^2}{(4\pi)^2.d_{tx}^2.L} \cdot \left(\frac{d_0}{d}\right)^2 = P_{rcvd}(d_0). \left(\frac{d_0}{d}\right)^2$$

Where, $P_{tx}$ is the transmission power, $G_t$ and $G_r$ are the **antenna gains** of transmitter and receiver. Where, $d_0$ is the so-called **far-field distance**, $\lambda$ is the **wavelength** and $L \geq 1$ summarizes losses.

$$P_{rcvd}(d) = P_{rcvd}(d_0) \left(\frac{d_0}{d}\right)^{\gamma}$$

Where, $\gamma$ is the path loss exponent called log distance.

## 4          Address Assignment Algorithm

The  assignment  algorithm  must  take the  existence  of  asymmetric links  into  account  when  node a has  successfully  executed  the  assignment algorithm,  it  can  pick  an  own  address. Instead of pick a random address  a select lowest  possible  non conflicting  address  .in a sense an  address  selection is greedy but  this approach ,lower  address  are preferred  and  occur more after  in the network lower  address  have  a  higher relative     frequency  and  we  have  non uniform address   distribution.

**Content based addressing:** Several   content   based naming systems have appeared in the directed diffusion routing. The   attribute type specifies the  kind  of  sensor  to which  the   interest    is  directed(temperature sensor).the  next attribute  threshold from  below  specifies,  where a   threshold   of  20c is  erosed  from below  square between    (0,0) and  (20,20).

Address   matching   algorithm:

```
Parameters:
Attribute sets A and B
// A corresponds to the interest, B to the data message
For each attribute a in A
where a.op is formal {
matched = false
foreach attribute b in B
where a.key == b.key and b.op is actual
{    if  b.val satisfies condition
              expressed by a.key and a.val
      then {
                      matched = true
                }
}
if (not matched )
then {
      return false      }
}
return true; // matching successful !
```

Example Simulation work for address matching algorithm:

temperature sensor
<type ,temperature ,IS >
<x-coordinate ,10,IS >
<y-coordinate ,10,IS >
interest message
type ,temperature ,EQ >

<threshold -from -below ,20,IS >
<x-coordinate ,20,LE >
<x-coordinate ,0,GE >
<y-coordinate ,20,LE >
<y-coordinate ,0,GE >
<interval ,0.05 ,IS >
<duration ,10,IS >
<class ,interest ,IS >

***Operator name:***
EQ →Matches if actual value is equal to value
NE →Matches if actual value is not equal to value
LT → Matches if actual value is smaller than value
GT →Matches if actual value is greater than value.

LE → Matches if actual value is smaller or equal to value
GE → Matches if actual value  is  larger  or  equal to value

# 5          Implementation of Key Method

Simple  key  distribution  controller   is  one  of  the  simplest  solution   for   key
management         a  group  controller  $C$  shares  a  secret  key  $keyC,i$   with  each
groupmember G$Mi$ . The groupcontroller is responsible for generating a group secret
key $kG,$. To distribute the group key, $C$ encrypts the group key with $kC,i$ and unicasts
it to $Mi$ . When a new member $MN$+1 joins the group, the group controller generates a
new group key $k'G'$,encrypts it with the old group key $kG$, and multicasts it to $\{M1,$
$M2, \cdots, Mn\}$.   The  group  controller  encrypts  the  new  group  key  $k'G'$  using
$kC,N$+1.

## 5.1 Key relationship Notations:

          **Triple(m,n,r) :**where **m** is a finite and nonempty set of group members, **n** is
a finite and nonempty set of keys, and $R \subset M \times N$ is a binary member-key relation.

Let assume *keyset(Me )* ,*user set(kn)*
*keyset(Me )* = $\{k|(Me, kn) \in R\}$ represents the set of all keys held by member $M$
*keyset(φ)* = *φ. user set(k)* = $\{Me \ |(M, k) \in R)\}$, represents the set of all members
holding key $k$.

## 5.2 Mathematical Implements:

          A key graph where,
                    *keyset(M2)* = $\{k12, k234, k1234\}$, *keyset(M3)* = $\{k23, k1234\}$,
                    *user set(k1)* = $\{M1\}$ and *user set(k12)* = $\{M1, M2\}$, where $ki \cdots j$

The following actions are performed by the group controller:

$$C \rightarrow \{M1, \cdots, M6\} : \{k1{-}9\}k1{-}8$$
$$C \rightarrow \{M7, M8\} : \{k1{-}9, k789\}k78$$
$$C \rightarrow M9 : \{k1{-}9, k789\}k9 .$$

The member-join, the group $\{M1, \cdots M8\}$, changes to $\{M1, \cdots, M9\}$ and subgroup $\{M7, M8\}$ changes to $\{M7, M8, M9\}$. $M1, \cdots M6$ belong to subgroups whose compositions are not affected by the new member-join, and, they only receive the new session key $k1{-}9$ encrypted using the old group key $k1{-}8$. Whereas the composition of the group containing $M7$ and $M8$ is affected by the new member-join.

When $M9$ leaves the group, the group controller performs the following actions:

$$C \rightarrow \{M1, M2, M3\} : \{k1{-}8\}k123$$
$$C \rightarrow \{M4, M5, M6\} : \{k1{-}8\}k456$$
$$C \rightarrow M7 : \{k1{-}8, k78\}k7$$
$$C \rightarrow M8 : \{k1{-}8, k78\}k8 .$$

When $M9$ leaves the group, $M1, \cdots M6$ cannot use the old group session key $kG = k1{-}9$ to encrypt the new session key $k\_G = k1{-}8$, because $M9$ knows theold key $kG$.

### 5.3  Modulated  signal

S(t)=A(t).cos(w(t)+φ(t))
  Where,
  A(t)=time  dependent  amplitude
  W(t)=time  dependent  frequency
  Φ(t)=phase  shift
  S(t)=time  dependent sensor node

$$S_i(t) = \frac{\sqrt{2ei(t)}}{t} . \cos [wt +]$$

Where $w_0$ is  the  centre  frequency  ,φ is  an  arbitrary  constant  initial phase .$E_i(t)$ is  constant  over  symbol  duration  (0,t). We  assume m different levels.

$\frac{\sqrt{2e}}{t}$ Explicitly  symbol  energy E.

Binary  data  string  110100101 is  modulated  using  $E_0(t)=0$,  $E_1(t)=1$. t-switching of  the  transmitter  is called  ON-OFF  Keying(OOK).

## 6    Conclusion

In this paper, we study the key methodology key code data exchange protocol, compression P – Box, Path loss attenuation, and address assignment algorithm, content based addressing and address matching algorithm, time synchronization modulation signal, mutual authentication, authentication key mechanism for secure data collection in wireless sensor network.